\documentstyle[preprint,aps,prbbib]{revtex}
\draft
\begin{document}
\title{Metal induced gap states and
Schottky barrier heights at non--reactive GaN/noble metal interfaces}

\author{S. Picozzi and A. Continenza}
\address{
Istituto Nazionale di Fisica della Materia (INFM)\\
Dipartimento di Fisica,
Universit\`a degli Studi di L'Aquila, 67010 Coppito (L'Aquila), Italy }
\author{G. Satta and S. Massidda}
\address{Istituto Nazionale di Fisica della Materia (INFM)  \\
Dipartimento di Scienze Fisiche \\ Universit\`a degli
Studi di Cagliari,  09124 Cagliari, Italy \\ and}
\author{A. J. Freeman}
\address{Department of Physics and Astronomy and Materials Research Center\\
Northwestern University, Evanston, IL 60208 (U.S.A.)\\}

\maketitle

\begin{abstract}
We present   ab-initio local density FLAPW calculations
on non--reactive N--terminated [001] ordered
 GaN/Ag and GaN/Au interfaces and compare the results (such as metal induced 
gap states  and Schottky barrier heights) 
 with those obtained for  GaN/Al, in order to 
understand the dependence of the relevant electronic properties on the 
deposited metal. Our results show that 
the density of gap states is appreciable only in the 
first semiconductor layer close to the interface. 
The decay length of the gap states in the semiconductor side
is  about 2.0 $\pm$ 0.1 \AA $\:$ and is independent
of the deposited metal, therefore being to a good extent
a bulk property of GaN. Our  calculated
values of the Schottky barrier heights are
$\Phi_{B_p}(GaN/Ag)$ = 0.87 eV and $\Phi_{B_p}(GaN/Au)$
= 1.08 eV; both values are smaller than the GaN/Al value 
($\Phi_{B_p}(GaN/Al)$ = 1.51 eV) and this quite large
spread of  values
 excludes the possibility of a Fermi level pinning within the GaN
band gap.
Because of the low screening  in GaN,
 the potential barrier at the junction
  is strongly affected by the structural arrangement of the first 
metal layer at the interface. This  leads to quite large
variations of the Schottky barrier height as a function of the metal,
in contrast with the behavior of GaAs/metal interfaces.
\end{abstract}


\section{Introduction}
GaN has certainly been one of the most studied compounds in 
the last few years, mainly because of both interesting optical properties and 
remarkable thermal stability, which render this semiconductor particularly 
suitable for important technological applications. As is well known, however, 
device performances depend on good metallic contacts and so  the study 
of Schottky barrier heights (SBH) in GaN/metal systems is of great
relevance: as an example, the performance of GaN--based laser diodes
is still limited by the difficulty in making low resistance ohmic contacts.
In this regard, surface reactivity and the presence of interface states are
also seen to play a relevant role in  Schottky barrier formation. 

In a previous work, we investigated the GaN/Al system, which is
considered to be a {\em reactive} interface due to the Ga--Al exchange
reaction driven by AlN formation at the immediate interface; we
studied the
ideal interface \cite{slv5}, as well as the effects on the 
interface properties of 
some defects (such as atomic swap and Ga$_x$Al$_{1-x}$N
intralayers
\cite{slv6}) at the initial stages of the SBH formation \cite{slv7}.

In the present work, we report results of ab-initio calculations for  GaN/M
interfaces (with M = Ag, Au)  which are considered to be 
non--reactive\cite{wu98} and compare the results (such as metal induced 
gap states (MIGS) and SBH)  
 with those obtained for  GaN/Al, in order to 
understand the dependence of the relevant electronic properties on the 
deposited metal. The interest in studying noble metal  contacts
resides in  understanding the effect of the $d$ states on the
Fermi level position, which has been thought to be relevant
in the case of GaAs interfaces \cite{schilf90}.
Moreover, we investigate the role of the atomic positions in the interface
region in determining the final SBH values: starting from GaN/Al,
line--ups differing by as much as 0.80 eV  can be produced
   by  changing
the interface N-metal
interplanar distance from its equilibrium value to that corresponding to
the GaN--Ag interface; as a result, the Schottky barrier height is brought to a
value  very close to that obtained for  GaN/Ag.
This leads to the conclusion that strain effects, mainly affecting the
magnitude of the interface dipole, play a major role in determining the final
SBH at the GaN/metal interface.

\section{Technical details}
The calculations were performed using the all--electron
full-potential linearized augmented plane wave (FLAPW)
\cite{FLAPW}  method
within density functional theory in the local density approximation (LDA).
We used a basis set of plane waves with wave vector up to $K_{max}$ = 3.9 a.u.,
leading to about 2200 basis functions and for the potential and the charge density
we used an angular momentum expansion up to $l_{max}$ = 6; tests performed
by increasing  $l_{max}$ up to 8 showed changes in the Schottky barrier
height of less than 0.03 eV.  
The Brillouin zone
sampling was performed using 10 special $k$--points  according to  the
Monkhorst-Pack scheme \cite{MP}.
 The muffin tin radii, $R_{MT}$, for Au and Ag were chosen equal
to 2.1 a.u., while for Ga and N
we used $R_{MT}$ = 1.96 and 1.65 a.u., respectively. 
We have considered supercells containing 15 GaN layers (8 N and 7 Ga atoms)
and 9 metal layers; tests performed on the cell dimensions have shown that
bulk conditions are well recovered far from the interface using this 15+9 layer
cell 
size (see discussion below).

\section{Structural properties}
 GaN is well known to show polytypism between the zincblende
and the wurtzite phase, so that either one can be easily stabilized;
we therefore concentrate on [001] ordered  N--terminated
zincblende interfaces in order to avoid the contribution of spontaneous 
polarization effects inside GaN that might  contribute to the Schottky barrier
height. Our goal is in fact to  investigate the
role played by the different metals in determining the position
of the Fermi level within the semiconductor band gap. 
We considered
 the metal as grown epitaxially on a GaN substrate ($a_{subs}$ = $a_{GaN} 
= 4.482$ \AA). 
Given the bulk lattice constants of the three metals, 
$a_{\rm Al} = 4.05$~\AA, $a_{\rm Ag} = 4.09$ \AA\ 
and $a_{\rm Au} = 4.08$ \AA, 
all the metals considered show a quite large mismatch with the GaN
substrate,
ranging from 8.8 \% in the case of fcc--Ag up to 9.6 \%
in the case of fcc--Al. In all cases, their  
lattice constants are smaller than that of the substrate, which 
 implies that appreciable bond length relaxations
are expected for the metal overlayers.
We calculated the most stable structures, assuming  pseudomorphic
growth conditions and a geometry in which the metal atoms simply
replace the Ga atoms on their fcc-sites, using total energy minimization
and the
{\it ab initio} forces calculated on each atomic site to find the
equilibrium values of  the interface
Ga-N, N-M and M-M interplanar distances.
In--plane relaxations, as well as the possibility 
of in-plane reconstruction of the GaN surface, 
before or during  metal deposition,  were neglected.
In this work, in fact, we are  interested mostly in studying the effect
of the metal overlayer on the interface GaN/M electronic
properties rather than determining 
 the structural configurations that may occur experimentally.

Our structural data are reported in Table~\ref{distances}.
Due to a refinement \cite{correzione} of our previous calculations on the 
GaN/Al system,
the  data  listed in Table~\ref{distances} for this system
differ  from those  already published\cite{slv5};
the largest change occurs for  the Al--Al bulk interplanar distance.
For clarity, we report our previous results in parenthesis in this same Table.
Let us focus on the structural rearrangement of 
Al, Ag and Au on the GaN substrate and consider
some relevant interplanar distances ({\em i.e.} distances between atomic 
planes along the [001] growth direction). 
A comparison between the free-electron-like case of Al 
 and the behavior of noble metals shows that none of the metals considered
seems to alter appreciably the bond length  at the
interface semiconductor layer so that the interplanar distance remains close
to its bulk value (1.12 \AA);  the larger deviation,
found in the case of Ag,  still  gives a reduction of the Ga--N interface 
distance by less than 5\%. 
A larger difference 
is found for the interface nitrogen--metal interplanar distance: 
the N--Al  distance is smaller 
than those corresponding to the Ag and Au structures by about 20~$\%$. 
This can be  related to the  different bonding at the interface in
the different cases - as will be further discussed  later on.
For both Al and Au we find that the  interplanar metal--metal distance,
$d^{M-M}_{int}$, increases compared to the bulk, leading to a bond--length
larger than equilibrium just at the interface layer. This effect is  far
more evident in the case of Al and can be related to a weakening of the
$s$--type metallic Al--Al bond due to a partial $s-p$ hybridization of the
interface aluminum in the Al--N covalent  bond.  We find
that already in  the sub--interface
layers  the forces are very small, thereby indicating that the metals
recover quickly their strained bulk--tetragonal bond--lengths.

In the last column of Table~\ref{distances}, we report the interplanar
distances calculated according to the macroscopic theory of elasticity
(MTE)\cite{vandewallemte} for the tetragonal metal strained to match the GaN
substrate, using  the
bulk elastic constants \cite{ashcroft} and the equilibrium bond lengths
\cite{harrison}
as input parameters. We recall that in all   cases considered, the mismatch
is pretty large (about 9\%) so that we might be probably out of the 
range of validity of the MTE. 
In fact, the discrepancies between 
the optimized interplanar distances within the
bulk regions (namely, $d^{M-M}_{bulk}$) and those predicted by MTE
range between 1\% in the case of Au to up to 8\% 
in the Al case. As expected, the bond length distances in the metal
bulk side are very similar for all the metals considered, seeing that their
equilibrium lattice constants are very close. 

\section{MIGS: the noble metal case and  comparison with the 
GaN/Al system}

We compare in Fig.\ref{pdosganau}  the atomic site--projected partial
density of  states  (PDOS) 
for the GaN/Au and GaN/Al  interfaces and inner N 
and M atoms,
taking the valence band maximum (VBM) of the inner N PDOS as the zero 
of the energy scale
(vertical  arrows denote the position of $E_F$).
The GaN/Ag PDOS is very similar to that of GaN/Au, and  are therefore
 not shown.
In order to  demonstrate the presence of the MIGS and the strong effect of
the metal
deposition on the interface semiconducting atoms, we show as  reference 
 the PDOS of the same atoms (N and metal) in the corresponding bulk
compounds ({\it i.e.} zincblende GaN and bulk metal). 
The PDOS for the interface N and Au
atoms - essentially due to $p$ and $d$ states, respectively
- (Fig.~\ref{pdosganau} (c) and (d))
show  peaks with a quite high density of states in the GaN band gap energy region
({\it i.e.}, between 0 and 1.8 eV). We recall here that LDA strongly
underestimates the GaN band gap  whose experimental value
is  $E_{gap}^{expt}$ = 3.39 eV \cite{foresi,bermudez});
 these states are seen to disappear
in the inner  bulk atom (Fig. \ref{pdosganau} (a) and (b)). 
Therefore, the presence of the metal 
affects only  the  semiconductor layers closer to the interface; already in
 the second layer (not shown) 
the MIGS decrease appreciably and the DOS gets very close to their bulk 
shape. Bulk conditions are  perfectly recovered in the inner layers
(see the
practically overlapping lines in Fig.~\ref{pdosganau} (a) and (e)), showing
that the  supercell dimensions are sufficient for our 
purposes. In particular, we notice that the LDA GaN gap is recovered
in the PDOS of the inner N atoms.

Apart from the presence of
MIGS, the  PDOS for both the N and Au  interface atoms shows
strong differences relative to the bulk, much larger than in the
GaN/Al case, where the most relevant difference consists in the general
modulation which brings the Al PDOS  from the free--electron
square--root--like 
behavior closer to the   PDOS of bulk AlN. The reason for this behavior
is that in  GaN/Au, the Au $3d$ 
states are occupied and  interact strongly with the N p states, which are 
also filled. As a result, the antibonding states rise in energy above the
semiconductor VBM and form the peaks at  around 1.5 eV, just in 
proximity to $E_F$.  
Such   features are 
absent in the GaN/Al case (Fig.~\ref{pdosganau} (g)) and are 
present in the PDOS of the interface N and Au  atoms
(Fig.~\ref{pdosganau} (c)),
completely disappearing on atoms far from the junction inside GaN.
The presence  of a peak at $E_F$ might indicate a tendency to an instability, 
probably leading to
in--plane reconstruction with the possible introduction of defects.
The spatial location of the charge density corresponding to the peak
around $E_F$
is shown in Fig.\ref{picco}: these states, which  
have a  clear anti--bonding character
between N and Au, are mainly localized in the
interface region with a resonant behavior inside the Au region (not shown)
and a negligible charge density in the GaN region. 
A similar situation occurs in the GaN/Ag system (not shown)
and was also reported for a [110] GaAs/Ag interface\cite{newman},
whereas it is completely absent in  GaN/Al-type structures due
to the lack of $d$--states in this  system. 

The overall shape of the
interface N PDOS shows  a depletion of states
in the region from $-4$ to $-1$ eV,
and a  peak around $-5$ eV, presumably  representing the  bonding partners 
of the  structures around $E_F$ and around the GaN VBM. This corresponds
to a degradation of the $sp^3$ bonding environment around the interface
N. On the basis of this discussion, 
 we can try to give an explanation for the
much larger distances $d^{N-Au}_{int}$ and $d^{N-Ag}_{int}$, compared with
 the Al case.  In fact,  the N-Al interface bond is similar to the
 one  in bulk AlN which provides its stability.
The N-Au bond, on the other hand,  mixes together filled states and 
pushes  towards higher energies
the antibonding 
combinations (with a large anion (N) contribution) around $E_F$ (but
 mostly below it).  This 
is consistent with the smaller amount of charge present inside  the 
interface N atom,
to be discussed later. In this scenario, decreasing the N-Au (Ag)
distances, would not    further stabilize the structure.

In order to investigate the spatial dispersion of the occupied gap 
states, we show in Fig.~\ref{migs}  the  macroscopic \cite{raffa} average
of the MIGS charge density in  GaN/Al (solid line), GaN/Au (dotted line)
and GaN/Ag (dot-dashed line). 
As already pointed out for GaN/Al  \cite{slv5}, 
 the presence of the metal 
affects almost exclusively  the interface semiconductor layer;  
the MIGS   decay exponentially,
approaching  zero   inside the semiconductor. 
Due to the different density of states distribution
of Al and Au (or Ag) for energies close
to $E_F$ and the consequent different positions of $E_F$ with respect to the
GaN VBM, the total integrated MIGS charge in the whole cell (metal side
included) is larger in the
free--electron metal--like case;  
 however,
the  behavior of these states as a function of the distance from the junction
in the three interfaces is overall very similar.
Actually, we find  from Fig. \ref{migs} a very similar  behavior,
which can be extrapolated with an exponential,
leading to the same   decay length  (see
Ref. \onlinecite{slv5} for details)  
for  both GaN/Au and GaN/Ag estimated to be  
 $\lambda$ = 2.0 $\pm$ 0.1 \AA., which is close to
the value obtained
for the GaN/Al system ($\lambda\:\approx$ 1.9 \AA~\cite{slv5}).
Therefore,
even though the energy dispersion of the MIGS  is somewhat different in the
free--electron--like and noble metal interfaces (see Fig. \ref{pdosganau}),
they are equally screened in the semiconductor side within 1--2 layers,
thus showing
that, within  a good approximation, $\lambda$ is a  GaN bulk property.

We plot in Fig. \ref{cori} the binding energies of $1s$ core levels, 
with respect to the $E_F$ 
 of Ga and N  (panels (a) and (b) respectively) 
  and  the difference between MT charges in the GaN/M
superlattices (SL) and in  GaN bulk  (Fig.\ref{cori} (c) and (d)
for the Ga and N atoms, respectively) as a function of the distance
from the interface.
As already discussed in Ref. \onlinecite{slv5}, it is clear that 
in the Al case the charge rearrangement at the 
interface causes a small effect (about 0.15 eV)
on the interface N core level
and negligible effects on the other semiconductor
atoms. On the other hand, in the noble metals case, the interface effect is 
much stronger:  the GaN interface layer shows  core level binding energies 
 differing 
by about 0.4  (0.5) eV for Ag (Au), from the values 
of the inner  bulk-like atoms. Moreover, the core levels
{\em bending}  
in going from the bulk
towards the interface  follows an opposite trend in the noble and
free-electron--like metals.
This  behaviour is consistent with the trend of the
valence charge inside  atomic
spheres: the interface nitrogen atom shows a
charge depletion (enhancement) with respect to the bulk atoms in the noble 
metal (Al) case. As  discussed above,  
the presence of a  peak (see Fig.\ref{pdosganau})
in the GaN gap energy region, with 
 antibonding N~$p-$Au~$d$ character, 
might be the cause of this  charge rearrangement and the resulting chemical
shift observed on the core level profiles.

\section{Schottky barrier heights}

To calculate the values of the SBH,  we
 adopt the usual procedure \cite{az,mmf} which takes  core
levels as reference energies. In particular, the potential discontinuity can
be expressed as the sum of two terms: $\Phi_B = \Delta\:b+\Delta\:E_b$, where
$\Delta\:b$ and $\Delta\:E_b$ denote an {\em interface} and {\em bulk} contribution,
respectively.
We evaluate $\Delta\:b$ taking the difference of Ga $1s$ and
the noble--metal $1s$ core levels energies in the superlattice:
$\Delta\:b= E_{1s}^{Ga}- E_{1s}^{NM}$. On the other hand, the {\em bulk}
contribution can be evaluated from separate  calculations for bulk GaN and
noble--metal and calculating the difference between the binding energies of the
same $1s$ levels considered above: 
$\Delta\:E_b = (E_{VBM}^{GaN}- E_{1s}^{Ga}) - 
(E_{F}^{NM}- E_{1s}^{NM})$. 

The $p$ type
SBH values obtained,  shown in Table~\ref{SBH},  include a
spin--orbit perturbation 
$\Delta_{SO}^{GaN} \: \approx$ 0.1 eV, but do not include quasi--particle
corrections.
Note that, as already  pointed out, this refinement
in the calculation of the GaN/Al system \cite{correzione}
gives
a $4 \%$ difference in the N-Al interface distance (which changes the
core level alignment at the interface) and 
a large difference in the tetragonal bulk Al
interplanar distance. This last quantity affects the Al core 
level binding energies used in the evaluation of the final SBH, 
 changing considerably the  
{\em bulk} contribution, 
essentially related to the absolute deformation potential
of the Fermi level in  bulk Al.  As a result,
we obtain a SBH value  different  from the one  previously published 
\cite{slv5} ($\Phi_B$ = 1.12 eV).
 The values shown in Table~\ref{SBH} are  in good agreement (within 0.1-0.2 eV)
 with those calculated from the density of states, obtained by
 considering 
the SBH as the energy distance between $E_F$
(see vertical arrows in Fig.~\ref{pdosganau}) and the top of the
valence band   of the PDOS corresponding to the inner semiconductor layer
inside the bulk region of the superlattice (the energy zero in 
Fig.\ref{pdosganau}). 

We note that the SBH values in the noble--metal case are  lower 
 than the SBH in  the GaN/Al interface
($\Phi_{B_p}(GaN/Al)$ = 1.51 eV).
To better understand
 this result, we should consider that the Al and noble metal
interfaces  differ in two main
aspects: ({\em i}) different chemical species
of the metal overlayer and  ({\em ii}) different
structural properties, $i.e.$ different bond lengths
at the interface  that comprise, in all respects, an interfacial  strain
contribution. In particular, from inspection of Table~\ref{distances} it
is evident that the   Ag and Au structures show very similar
$d^{N-M}_{int}$ and $d^{M-M}_{int}$ interplanar distances, which are
at variance with
those with Al.  
In order to separate the chemical from the strain contribution, we evaluate
the SBH for three different structures that can be regarded as intermediate
steps necessary to bring the GaN/Al structure to  match perfectly the
GaN/Ag one. 
We show in Table \ref{step} the interface interplanar distances and
final SBH values for the
equilibrium GaN/Al and GaN/Ag systems and for the three intermediate
interfaces.  
In the first structure (Step I), the interplanar $d^{N-M}_{int}$ distance 
of the GaN/Al SL is taken equal to that minimized for the GaN/Ag SL
($d^{N-M}_{int}$ = 1.32 \AA) : the SBH is
reduced from 1.5 eV to 0.76 eV (this surprising  result will be 
discussed in  detail later on).
 
 As a second step (Step  II), we  change  
$d^{M-M}_{int}$ to recover that calculated for the GaN/Ag structure:
$d^{M-M}_{int}$ = 1.60 \AA. 
The SBH is remarkably less sensitive to this parameter,
giving only a 0.04 eV change in the potential barrier which
brings the SBH to about 0.80 eV. This 
is expected since the metal efficiently
screens out  the perturbation generated by  atomic
displacements: actually, 
the dynamical effective charge, related to the dipole induced
by a unit displacement of atoms, is zero inside a metal. 

In the third 
 structure (Step III),  
 the Ga-N interface  distance is brought to its value in the 
GaN/Ag superlattice,  $d^{Ga-N}_{int}$ = 1.07 \AA,
and we have a system where Al atoms  perfectly replace Ag 
in GaN/Ag. Although this last structural
 change is very small (about 4 $\%$), this interplanar
 distance  turns out to be very important for the final potential line--up,
due to  the incomplete screening in the semiconductor
 side and the large N effective charge.  As a result,  
 the SBH changes appreciably ($\Phi_B$ = 0.71 eV).
 This  final result is quite  close (within 0.2 eV) to the
value found for the real GaN/Ag interface,  showing that apparently
the interface strain plays a more important role than the bare
chemical contribution.

Perhaps the most surprising result of our tests is the very strong
dependence of the SBH on the interface N--Al distance, whose variation
represents the larger contribution to the difference  between GaN/Al and
GaN/Ag (Au) SBHs. Test calculations have shown an almost perfect linear
behaviour of $\Phi_B$ against $d^{N-Al}_{int}$, leading to an Al 
effective charge $Z^*_L=0.08$, ($Z^*_L=Z^*_T/\epsilon_{\infty}$, 
where $Z^*_T$
is the Born dynamical charge and $\epsilon_{\infty}$ is the electronic
static dielectric constant).
This result is in sharp contrast with  the case of 
GaAs/Al
\cite{alice}, where no significant changes were found for small elongations
of the As-metal interface distance, therefore resulting in
 $Z^*_{L}\sim$ 0 - an almost perfect metallic behavior.
 As a further test, 
we performed  calculations on  GaAs/Al, similar to those of Ruini
{\it et al.}\cite{alice}, confirming their results both in terms of  the 
SBH values, and of their behavior as a function of  $d^{As-Al}_{int}$.

A possible  explanation of the striking
difference between the GaN/Al and GaAs/Al systems can be provided by an 
analysis of the relevant physical parameters involved. 
 The MIGS 
decay length is larger for GaAs/Al ($\lambda_{GaN}\approx$ 2 \AA\  and 
$\lambda_{GaAs}\approx$ 3 \AA\cite{louie,baldi}), therefore 
providing a more extended  region with metallic
behavior inside the semiconductor; in addition, the density of states at 
the Fermi
level, $N(E_F)$, is  larger for GaAs/Al (see Fig~\ref{dosef}), giving rise 
to a  smaller (by a factor of $\approx 1.5$) Thomas-Fermi screening 
length $\lambda_{TF}$ in this system.
Still, the  values of the MIGS decay length 
and of  $N(E_F)$ are not {\it very} different between GaAs and GaN, so 
that it is not at all obvious to expect such a behavior   
of the  SBH in the two compounds. 

To understand these results, we can use elementary electrostatics
arguments. If we make a few rough assumptions such as 
a simple Yukawa--like screened
potential and consider  a metallic behavior inside the
semiconductor up to distances of the order of
$\lambda$, the MIGS decay length, we find
that the potential difference across the metal/semiconductor junction
induced by  displacements of the Al interface atom scales with the
factor $e^{-k_{TF}\; \lambda}$.
If we now estimate the values of the Thomas--Fermi screening lengths, $k_{TF}$,
using the GaN/Al and GaAs/Al superlattice  value of
$N(E_F)$, we are led to the conclusion that a unit displacement
of the interface Al atoms produces a potential change across the
interface which is roughly 7 times larger in GaN/Al than in GaAs/Al.
Considering the crudeness  of this model,  such an estimate is in 
 satisfactory 
agreement with the first--principles results, which indicate a factor of
$\approx 10$ for the same ratio. In other words, GaAs screens out
almost perfectly all the structural changes in the interface
region (namely,  displacements of the interface Al
atoms),
while the same is not true for GaN.

In order to further investigate  the effects of the $d^{N-M}_{int}$
change, we compare the core levels and MT charges for the equilibrium
GaN/Al interface and  the   Step I system. 
Recall that these systems
are exactly equal, except for the difference
in the interface nitrogen--metal distance: $d^{N-M}_{int}$ = 1.11 \AA
$\:$ and $d^{N-M}_{int}$ = 1.32 \AA, for the equilibrium and Step I
structures, respectively.
As in Fig. \ref{cori},
we show in Fig. \ref{cfr_eq_step} the trends for the
core level binding energies  (panels (a) and
 (b)  for Ga and N atoms,
respectively) and the difference between the MT charges
compared to their  values in the bulk compounds
(panels (c) and (d) for Ga and  N  atoms,
respectively), as a function
of the atomic distance from the interface. In terms
of core levels 
and charge transfer within the MT, a comparison with Fig. \ref{cori}
shows that the    Step I system behaves
very closely   the GaN/Ag structure but
very differently from the equilibrium GaN/Al interface.
Therefore,  the electronic charge distribution around
the interface N and Al atoms has a strong dependence on the
interplanar distance and is able to reduce the SBH
by as much as 0.75 eV, 
bringing a GaN/Al Step I SBH within only 0.16 eV from
the GaN/Ag SBH.

The dispersion of the SBH values seems to exclude a {\em Fermi level pinning}
in the GaN case, as experimentally confirmed by the large spread of values 
reported in the literature for the SBH between GaN and different metals.  
In particular, let us recall  those obtained for $n$-GaN/Ag and
$n$-GaN/Au: $\Phi_{B_p}^{expt}(n-GaN/Ag)$ = 2.7 eV \cite{ganag} and
$\Phi_{B_p}^{expt}(n-GaN/Au)\:\approx$  2.4 eV \cite{ganau1} and 2.2 eV 
\cite{ganau2,wu_khan}. 
On the other hand, recent photoemission measurements \cite{wu_khan}
performed for Au deposited on $p$-type GaN
show that the Fermi level is stabilized around 1 eV above the $p$-GaN VBM,
in apparent good agreement with our calculated value ($\Phi_B$ = 1.08 eV).
The disagreement
between some of these values and our calculated ones are certainly
related to the different conditions of the GaN surface (which is ideal
in our calculations and subject to different preparations in the 
experimental case); moreover, we considered the GaN zincblende structure
and   [001] oriented interfaces, whereas all the
experimental samples are grown on [0001] wurtzite GaN.

We  note finally that the spread of experimental values  
for GaN/metal interfaces has not been found for
  the GaAs/metal systems, where a Fermi level
pinning was experimentally observed \cite{gaasexp}. 
This  can be  understood in terms of
the much more efficient  screening provided by GaAs with respect to GaN, 
as discussed above. In
fact,  the main difference between the noble metal
and Al systems is the length of the anion--metal bond which gives rise to large
variations of the interface dipole: this electrostatic contribution is  
poorly  screened in GaN   and therefore contributes considerably
 to the final
SBH value. In addition, it has to be observed that GaAs matches better 
(within 2$\%$) the
Au, Ag and Al lattice constants so that the interface strain does
not play a crucial role in this case.

\section{Conclusions}

We have performed FLAPW calculations for  [001] ordered
GaN/Ag and GaN/Au interfaces, mainly focusing on the electronic
properties and comparing our results with those
obtained from previous
calculations  for the GaN/Al interface. 
Our calculations show that there is an appreciable
density of MIGS in the 
noble metal interfaces considered
(even higher than in the GaN/Al case); 
however, the presence of the gap states
is relevant in the interface layer only, being strongly reduced
already in the sub--interface layer. We 
estimate the  MIGS decay length  to be $\lambda\:\approx$ 2.0 \AA,
for all the metals considered (Al, Ag and Au).
The SBH values ($\Phi_{B_p}(GaN/Ag)$ = 0.87 eV and $\Phi_{B_p}(GaN/Au)$
= 1.08 eV) are significantly
 smaller than the value obtained in the GaN/Al case
($\Phi_{B_p}(GaN/Al)$ = 1.51 eV); we demonstrated 
that the appreciable SBH reduction
in going from the free--electron to the noble--metal case 
is  mostly due to  structural effects. In particular, the distance
between the last N  and the first
metal layer plays a critical role in dictating  
 the final SBH value, in contrast
with that found in GaAs/Al, where previous \cite{alice},  as well as our
present,
calculations showed   negligible effects of this same
structural parameter.  We found that the largest structural differences between
the various GaN/M interfaces considered 
are related to this distance, mainly
determined
by the different bonding nature between N and free--electron--like or noble
metals. 
Finally, we were able to show, at least for our perfectly ordered abrupt
interfaces, that the lack of Fermi level pinning in GaN can be understood in
terms of  electrostatic effects related to variations of the interface
anion--metal dipole: these effects
are not properly screened in GaN, so that they
 contribute considerably to the final potential line--up at the interface.

\section{ACKNOWLEDGEMENTS}
We gratefully acknowledge useful discussions with Prof. R. Resta and Dr.
A. Ruini.
Work in L'Aquila and Cagliari supported by grants
of computer time
at the CINECA supercomputing center (Bologna, Italy) through the Istituto
Nazionale
di Fisica della Materia (INFM).
Work at Northwestern University was supported by the U.S. National Science
Foundation through the Northwestern Materials Research Center.

\begin{table}
\centering
\small
\caption{Interplanar distances (in \AA) between the different atomic
planes in the GaN/Al, GaN/Ag and GaN/Au interfaces. 
 Values in parenthesis are taken from Ref. 1.}
\vspace{5mm}
\begin{tabular}{|c|c|c|c|c|c|}
 & $d^{Ga-N}_{int}$ &  $d^{N-M}_{int}$ & $d^{M-M}_{int}$& $d^{M-M}_{bulk}$ & 
 $d^{M-M}_{MTE}$\\
 \hline \hline
 GaN/Al & 1.11 (1.12)& 1.11 (1.15)& 1.88 (1.94)& 1.64 (1.94)& 1.78\\ 
 GaN/Au & 1.10 & 1.34 & 1.77 & 1.68 & 1.70 \\
  GaN/Ag & 1.07 & 1.32 & 1.60 & 1.64 & 1.75\\ 
\end{tabular}
\label{distances}
\end{table}

\begin{table}
\centering
\small
\caption{Schottky barriers (in eV)
at the GaN/Al, GaN/Ag and GaN/Au interfaces.}
\vspace{5mm}
\begin{tabular}[h]{|p{0.8cm}|p{3.5cm}|p{3.5cm}|p{3.5cm}|}
&GaN/Al & GaN/Au & GaN/Ag \\
 \hline \hline
$\Phi_B$ & 1.51 & 1.08 & 0.87\\
\end{tabular}
\label{SBH}
\end{table}

\begin{table}
\centering
\small
\caption{Interplanar distances (in \AA) between the different atomic
planes in the GaN/Al, GaN/Ag and GaN/Au interfaces (first three columns)
and Schottky barrier heights (in eV -- last column).}
\vspace{5mm}
\begin{tabular}{|p{2.cm}|p{2.5cm}|p{2.5cm}|p{2.5cm}|p{1.3cm}|}
 & $d^{Ga-N}_{int}$ &  $d^{N-M}_{int}$ & $d^{M-M}_{int}$& $\Phi_B$ \\
 \hline \hline
 GaN/Al & 1.11& 1.11 & 1.88& 1.51 \\ 
Step I & 1.11 & 1.32 & 1.88 & 0.76\\
Step II & 1.11 & 1.32 & 1.60 & 0.80\\
Step III & 1.07 & 1.32 & 1.60 & 0.71\\
  GaN/Ag & 1.07 & 1.32 & 1.60 & 0.87\\ 
\end{tabular}
\label{step}
\end{table}

\begin{figure}
\caption{ Inner (panels (a) and (e)) and interface (panels (c) and (g)) 
N atom PDOS in the GaN/Au  and GaN/Al systems, respectively. 
Panels (b) ((f)) and (d) ((h))
show the PDOS for the inner and interface Au (Al) atoms, respectively.
In all  panels, dashed lines indicate the corresponding atomic contribution
in the bulk materials.
The GaN VBM is taken as zero of the energy scale. Vertical arrows indicate 
$E_F$   with respect to the semiconductor VBM.}
\label{pdosganau}
\end{figure}

\begin{figure} 
\caption{3D plot
of the 
GaN/Au MIGS charge density due to the DOS peak in proximity to $E_F$
 projected on a plane which cuts the   N--Au
interface bond. Values on the $z$-axis in electrons/cell.}
\label{picco}
\end{figure}

\begin{figure}
\caption{Planar macroscopic
average of the MIGS charge density for  the GaN/Al (solid line),
GaN/Au (dotted line) and GaN/Ag (dot--dashed line) interfaces.}
\label{migs}
\end{figure}

\begin{figure}
\caption{Panels (a) and (b):
 core energy levels of Ga  and N  atoms, respectively,
for the GaN/Ag (stars), GaN/Au (squares) 
and GaN/Al (circles) systems as a function of the distance
from the interface. Binding energies
(in eV) are referred to the Fermi levels in the different systems
and were shifted by 10190.74 eV and 353.75 eV in the Ga and N case, 
respectively.
Panels (c) and (d): difference between MT charges in  GaN/metal superlattices (SL) and
in GaN bulk for Ga  and N  atoms, respectively,
 as a function of the distance
from the interface. Symbols as in panels (a) and (b).}
\label{cori}
\end{figure}

\begin{figure}
\caption{Total DOS of GaN/Al (solid line)  and of GaAs/Al (dotted line) in
proximity to $E_F$ (taken as zero of the energy scale). Vertical
arrows
denote the position of the semiconductor VBM.}
\label{dosef}
\end{figure}

\begin{figure}
\caption{Panels (a) and (b):
 core energy levels of Ga  and N  atoms, respectively,
for the  GaN/Al equilibrium (circles) and Step I (diamonds)
 systems as a function of  distance
from the interface. Binding energies
(in eV) are referred to the $E_F$.
Panels (c) and (d): difference between MT charges in  GaN/metal superlattices 
(SL) and
in GaN bulk for Ga  and N  atoms, respectively,
 as a function of the distance
from the interface. Symbols as in panels (a) and (b).}
\label{cfr_eq_step}
\end{figure}


\begin{thebibliography}{100}
\bibitem{correzione} We used a 75 special $k$-points sampling  
in the bulk tetragonal Al strained on GaN for the total
energy minimization of the Al-Al interplanar distance and a
10 special $k$-points sampling for the GaN/Al 15+9 supercell.
\end{thebibliography}
\end{document}